\documentclass{IEEEtran}

\usepackage{cite}
\usepackage{amsmath,amssymb,amsfonts}
\usepackage{graphicx}
\usepackage{textcomp,nicefrac}
\usepackage{upgreek}
\usepackage{hyperref}

\def\BibTeX{{\rm B\kern-.05em{\sc i\kern-.025em b}\kern-.08em
T\kern-.1667em\lower.7ex\hbox{E}\kern-.125emX}}

\begin{document}

\title{A two-prong Approach to the Simulation of DC-RSD: TCAD and Spice}

\author{T. Croci, L. Menzio, R. Arcidiacono, M. Arneodo, P. Asenov, N. Cartiglia, M. Ferrero, A. Fondacci, V. Monaco, A. Morozzi, \IEEEmembership{Member, IEEE}, F. Moscatelli, R. Mulargia, E. Robutti, V. Sola and D. Passeri, \IEEEmembership{Senior Member, IEEE}}

\thanks{Manuscript received July 19, 2023; revised July 19, 2023. This project has received funding from the European Union's Horizon 2020 research and innovation programme under GA No 101004761 and from the Italian MIUR PRIN under GA No 2017L2XKTJ.}
\thanks{T. Croci and A. Morozzi are with INFN, Section of Perugia, via A. Pascoli 23c, 06123 Perugia, Italy (e-mail: tommaso.croci@pg.infn.it).}
\thanks{L. Menzio, V. Sola, R. Mulargia and V. Monaco are with University of Torino, Department of Physics and with INFN, Section of Torino, via P. Giuria 1, 10125 Torino, Italy.}
\thanks{R. Arcidiacono and M. Arneodo are with Università del Piemonte Orientale, Largo Donegani 2/3, 20100 Novara, Italy and with INFN, Section of Torino, via P. Giuria 1, 10125 Torino, Italy.}
\thanks{F. Moscatelli and P. Asenov are with CNR-IOM of Perugia and with INFN, Section of Perugia, via A. Pascoli 23c, 06123 Perugia, Italy.}
\thanks{N. Cartiglia and M. Ferrero are with INFN, Section of Torino, via P. Giuria 1, 10125 Torino, Italy.}
\thanks{D. Passeri and A. Fondacci are with University of Perugia, Department of Engineering, via G. Duranti 93, 06125 Perugia, Italy and with INFN, Section of Perugia, via A. Pascoli 23c, 06123 Perugia, Italy.}
\thanks{E. Robutti is with INFN, Section of Genova, via Dodecaneso 33, 16146 Genova, Italy.}

\maketitle

\begin{abstract}
The DC-coupled Resistive Silicon Detectors (DC-RSD) are the evolution of the AC-coupled RSD (RSD) design, both based on the Low-Gain Avalanche Diode (LGAD) technology. The DC-RSD design concept intends to address a few known issues present in RSDs (e.g., baseline fluctuation, long tail-bipolar signals) while maintaining their advantages (e.g., signal spreading, $100\%$ fill factor). The simulation of DC-RSD presents several unique challenges linked to the complex nature of its design and the large pixel size. The defining feature of DC-RSD, charge sharing over distances that can be as large as a millimeter, represents a formidable challenge for Technology-CAD (TCAD), the standard simulation tool. To circumvent this problem, we have developed a mixed-mode approach to the DC-RSD simulation, which exploits a combination of two simulation tools: TCAD and Spice. Thanks to this hybrid approach, it has been possible to demonstrate that, according to the simulation, the key features of the RSD, excellent timing and spatial resolutions (few tens of picoseconds and few microns), are maintained in the DC-RSD design. In this work, we present the developed models and methodology, mainly showing the results of device-level numerical simulation, which have been obtained with the state-of-the-art Synopsys Sentaurus TCAD suite of tools. Such results will provide all the necessary information for the first batch of DC-RSD produced by Fondazione Bruno Kessler (FBK) foundry in Trento, Italy.

\end{abstract}


\section{Introduction}
\label{sec:introduction}

\begin{figure}[t]
\centerline{\includegraphics[width=3.4in]
%{images/FromACtoDC_RSD_EDITED_NEW2.PNG}}
{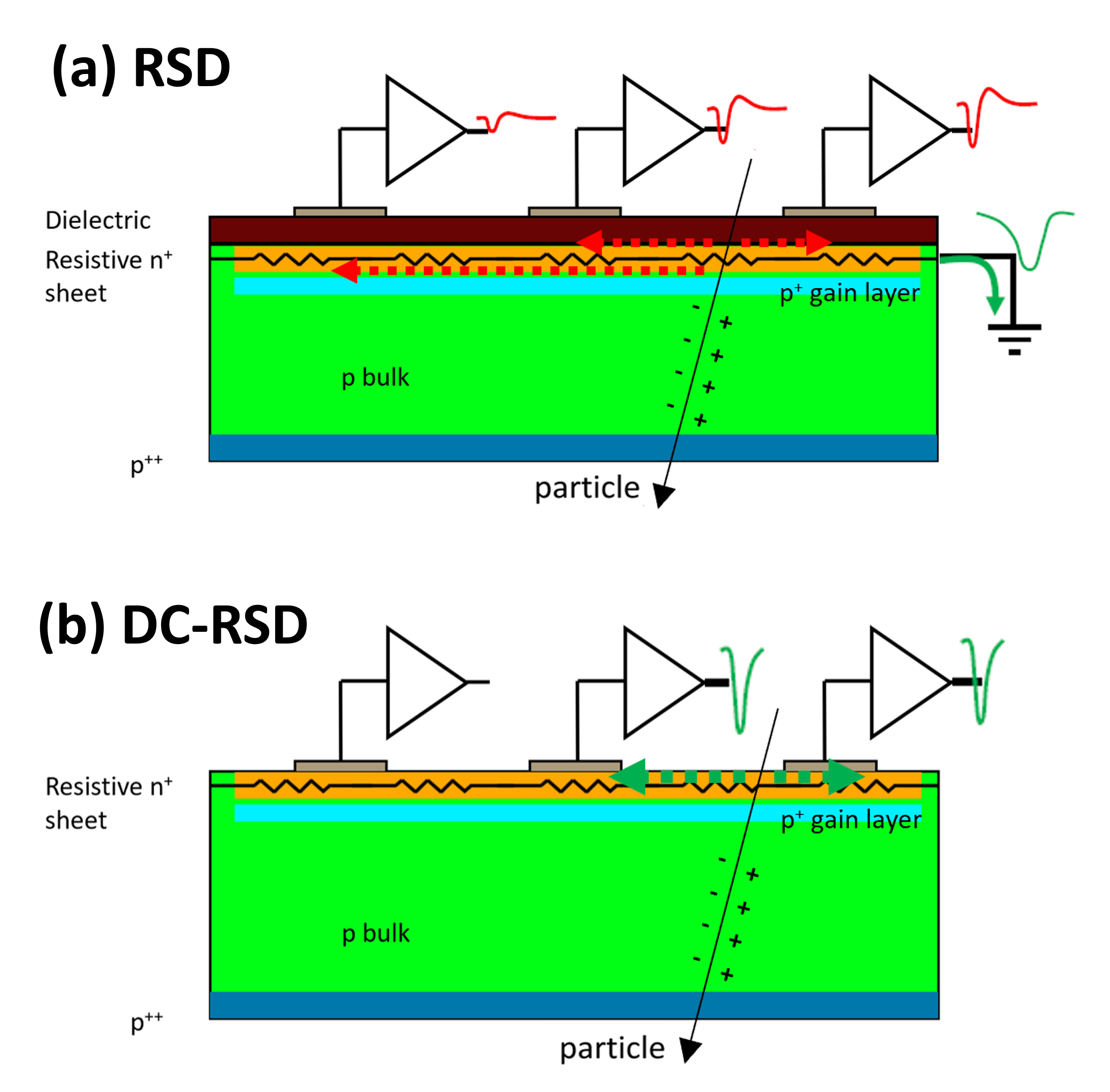}}
\caption{Schematic cross-section (not to scale) of an RSD (a) and a DC-RSD (b) sensor, both based on the LGAD technology. The $n^{+}$-resistive sheet and the $p^{+}$-gain layer implants are without any segmentation, which allows for signal sharing among neighbouring pads and a $100\%$ fill factor. In the DC-RSD design, the direct contact between the read-out pads and the $n^{+}$-resistive sheet leads to a better signal confinement with respect to the one of the RSDs.}
\label{fig1}
\end{figure}

\IEEEPARstart{I}{N} recent years, two innovative paradigms have been introduced in the design of silicon sensors for particle detection: the internal low-gain and the resistive read-out.
The internal low-gain has been introduced with the Low-Gain Avalanche Diode (LGAD) detector, which is a planar n-in-p silicon sensor that provides large signals and low noise, yielding a temporal precision in the order of tens of picoseconds \cite{a1}\cite{a2}. For this reason, the LGAD sensors are ideal to perform timing in particle detection. On the other side, the detector segmentation strongly affects the fill factor, resulting in low spatial resolution.
Thanks to the introduction of the resistive read-out in silicon sensors, it has been possible to avoid the detector segmentation, thus improving the spatial precision. The resistive read-out has been introduced with the AC-coupled Resistive Silicon Detector (RSD), that is essentially an LGAD-based silicon sensor optimized for both timing and spatial particle detection, which makes use of an AC-coupled read-out \cite{b1}\cite{b2}. As shown in \figurename~\ref{fig1}a, the major peculiarity of the RSD paradigm is the use of a continuous $n^{+}$-resistive sheet and a continuous $p^{+}$-gain layer that ensure signal sharing among neighbouring pads and $100\%$ fill factor, yielding excellent spatial precision with the usual temporal resolution of the LGADs \cite{b3}\cite{b4}. On the other hand, different drawbacks are linked to the nature of the RSD paradigm, which are \emph{(i)} the bipolar behaviour of the generated signals due to the use of the AC-coupled read-out, \emph{(ii)} the baseline fluctuation due to the collection of the leakage current only at the edge of the sensor, \emph{(iii)} a position-dependent resolution, and \emph{(iv)} the difficult to scale towards large area sensors.

We propose a new design of the LGAD-based resistive read-out sensors, that we called DC-coupled RSD (DC-RSD) \cite{b5}. As shown in \figurename~\ref{fig1}b, by eliminating the dielectric, and hence by performing a DC-coupled read-out, it is possible to have unipolar signals, absence of baseline fluctuation, a well-controlled charge sharing, and large sensitive areas (in the order of centimeters), while maintaining the advantages of signal spreading and $100\%$ fill factor.

%\quad

\section{Simulation of DC-RSD devices}
\label{sec:simulation}
The simulation of DC-RSD presents several unique challenges linked to the complex nature of its design and the large pixel size. The intrinsic feature of DC-RSD, i.e. the charge sharing over distances that can be as large as a millimeter, represents a formidable challenge for Technology-CAD (TCAD) device-level simulation tools, which ensure high accuracy and predictive capabilities at the expense of high computational costs. To circumvent this problem, we have developed a hybrid approach for the simulation of DC-RSD, which exploits a combination of two simulation tools at different abstraction level (circuit and device), namely Spice and TCAD.

Spice simulator ensures fast simulations at a higher abstraction level - the circuital one, so it is good for a proof of principle of the detector operation, but it delivers limited information. On the other side, TCAD allows to evaluate different technology options (e.g., the resistivity of the $n^{+}$ layer, contact materials) and geometrical layouts (shape and distance of the read-out pads), to model the injected stimulus and its trajectory, as well as to take into account the damage induced by the radiation both within the silicon bulk and at the silicon-oxide interface. In particular, a full 3D simulation domain guarantees an accurate evaluation of the electrical behavior and precise timing information, gaining access to the response of the detector device directly from its electrical contacts, overcoming the bandwidth limitation imposed by the read-out circuitry.

\subsection{Spice simulation}
\label{subsec:spice}
\begin{figure}[t]
\centerline{\includegraphics[width=3.4in]
% {images/LumpedElementElectricalModel.png}}
{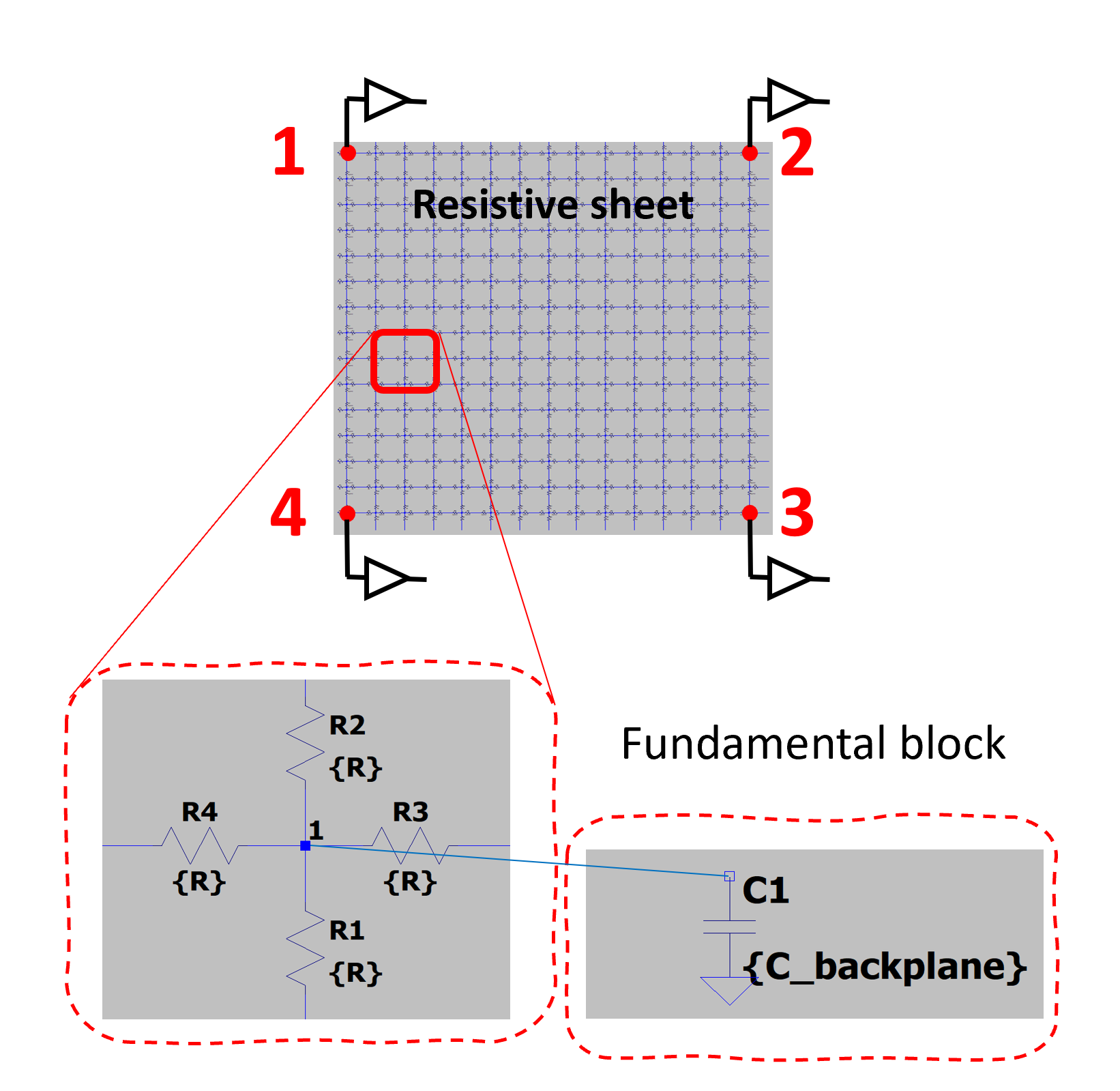}}
\caption{Schematic implementation in Spice simulator of a $340~\text{$\upmu$m}$-pitch four-pad DC-RSD device as a $15 \times 15$ nodes geometry. The combination of four resistors and a capacitor connected to the ground is the fundamental block of the full sensor model. The numbered red dots are the four pads of the sensor, each of which has been connected to the front-end electronics, whose input impedance has been modeled as a simple resistor.}
\label{fig2}
\end{figure}

In Spice environment \cite{c1}, we have developed the equivalent lumped-element electrical model depicted in \figurename~\ref{fig2}. The fundamental block of such model is a node connected to four resistors and a capacitor connected to the ground, which model the resistive sheet and the sensor backplane, respectively (see the inset in \figurename~\ref{fig2}). Multiple blocks connected to one another compose the full sensor electrical model. 

By injecting a test input signal in each node of the resistive grid, the generated current spreads over the resistive sheet and it is collected at the four corners of the full sensor electrical model - i.e., the four red dots in the schematic view of \figurename~\ref{fig2}. Thanks to this setup we have simulated the output waveforms in different conditions (e.g., in terms of sheet resistance and pitch size) with very short simulation times. 

\begin{figure}[t]
\centerline{\includegraphics[width=3.5in]
{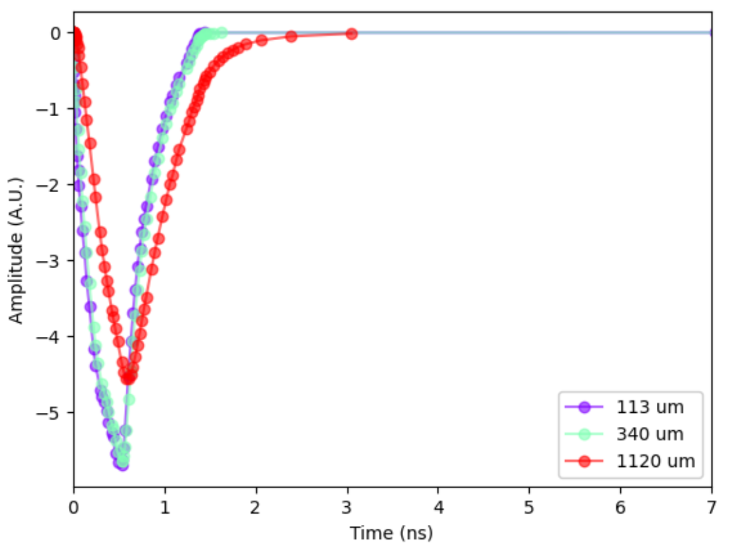}}
\caption{Variation of the signal shape simulated in Spice environment for different sizes of the sensor pitch - signal picked up from the top left pad of the lumped-element electrical model (picture taken from \cite{c2}).}
\label{fig2a}
\end{figure}

\begin{figure}[t]
\centerline{\includegraphics[width=3.4in]
{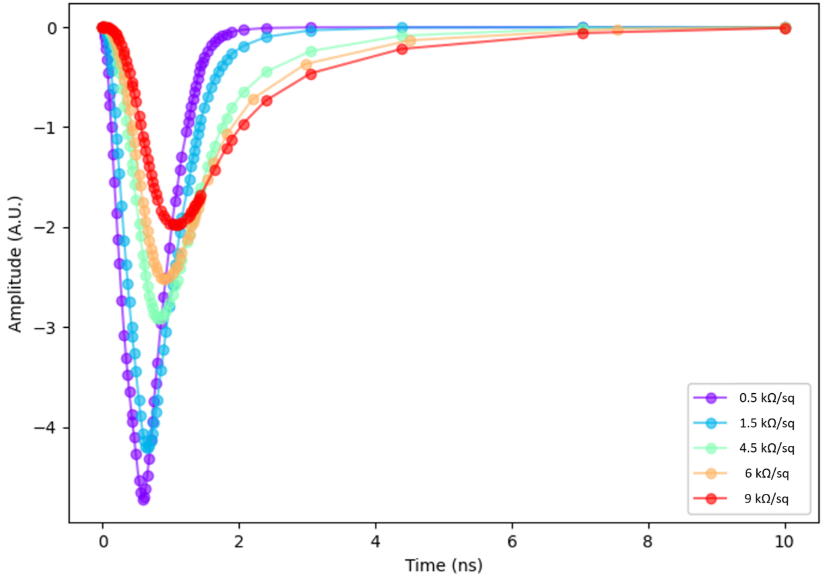}}
\caption{Variation of the signal shape simulated in Spice environment for different values of the sheet resistance - signal picked up from the top left pad of the lumped-element electrical model (picture taken from \cite{c2})}
\label{fig2b}
\end{figure}

In particular, by varying the backplane capacitance value, which is the only variable correlated to how large the sensor is, we observed that the generated currents maintain a sharp rising edge even for rather large pitch size - in the order of a millimetre (see \figurename~\ref{fig2a}). Moreover, an improvement of the timing capabilities has been observed by reducing the sheet resistance value so that the amplitude of the read-out signals becomes higher, and their duration shorter (see \figurename~\ref{fig2b}).

\begin{figure}[t]
\centerline{\includegraphics[width=3.38in]
% {images/Spice_Sim_PosRecon.png}}
{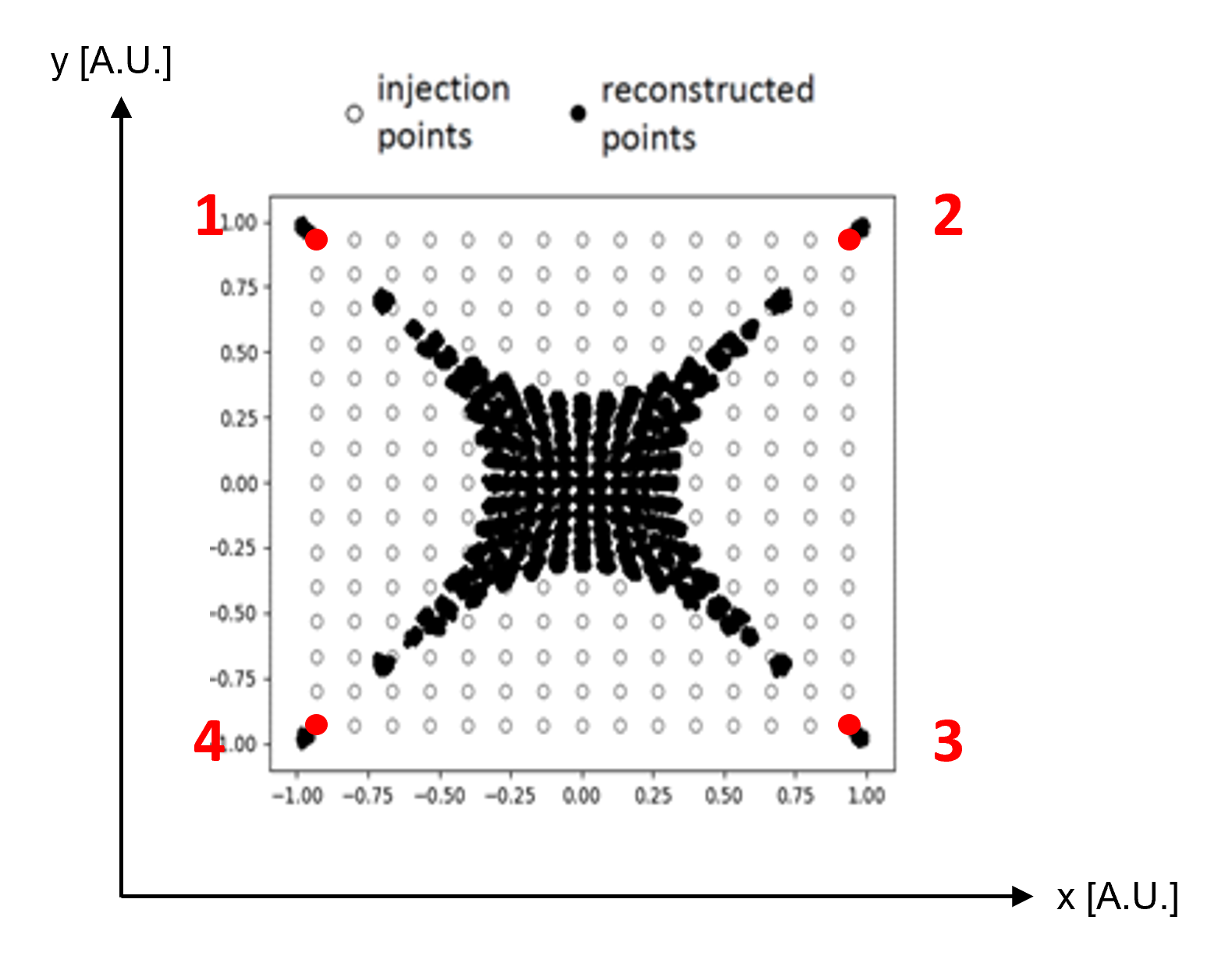}}
\caption{Map of the injected and reconstructed impact positions, which has been obtained by Spice simulations. The empty circles represent the injection points, and the filled circles are the reconstructed points.}
\label{fig2c}
\end{figure}

The spatial accuracy of the DC-RSD has been evaluated by comparing the injection positions of the test input signal with the relative reconstructed positions, as represented in the map of \figurename~\ref{fig2c}. The x-y coordinates of the reconstructed injection positions have been calculated by combining the amplitude of the signals collected by pairs of pads along opposite sides (top-bottom for the y-coordinate and right-left for the x-coordinate), according to the \textquotedblleft amplitude imbalance formula\textquotedblright:
\begin{equation}\begin{split}
\label{eq1}
x&=\frac{A_{2}+A_{3}-A_{1}-A_{4}}{A_{tot}}\\
y&=\frac{A_{2}+A_{1}-A_{4}-A_{3}}{A_{tot}}, 
\end{split}\end{equation}
where $A_{i}$ is the maximum amplitude of the signal read out by the $i$-th pad, and $A_{tot}$ the sum of the maximum amplitudes of the signal read out by the four pads \cite{c2}\cite{c3}.
By looking at \figurename~\ref{fig2c}, it is noticeable how the reconstructed positions do not overlap with the injected ones. A solution to this well-known problem (see section~\ref{sec:resolution}) has been explored during the detector design and optimization phases, and it will be presented in section~\ref{sec:optimization}.

\subsection{TCAD simulation}
\label{subsec:tcad}
\begin{figure}[t]
\centerline{\includegraphics[width=2.35in]
{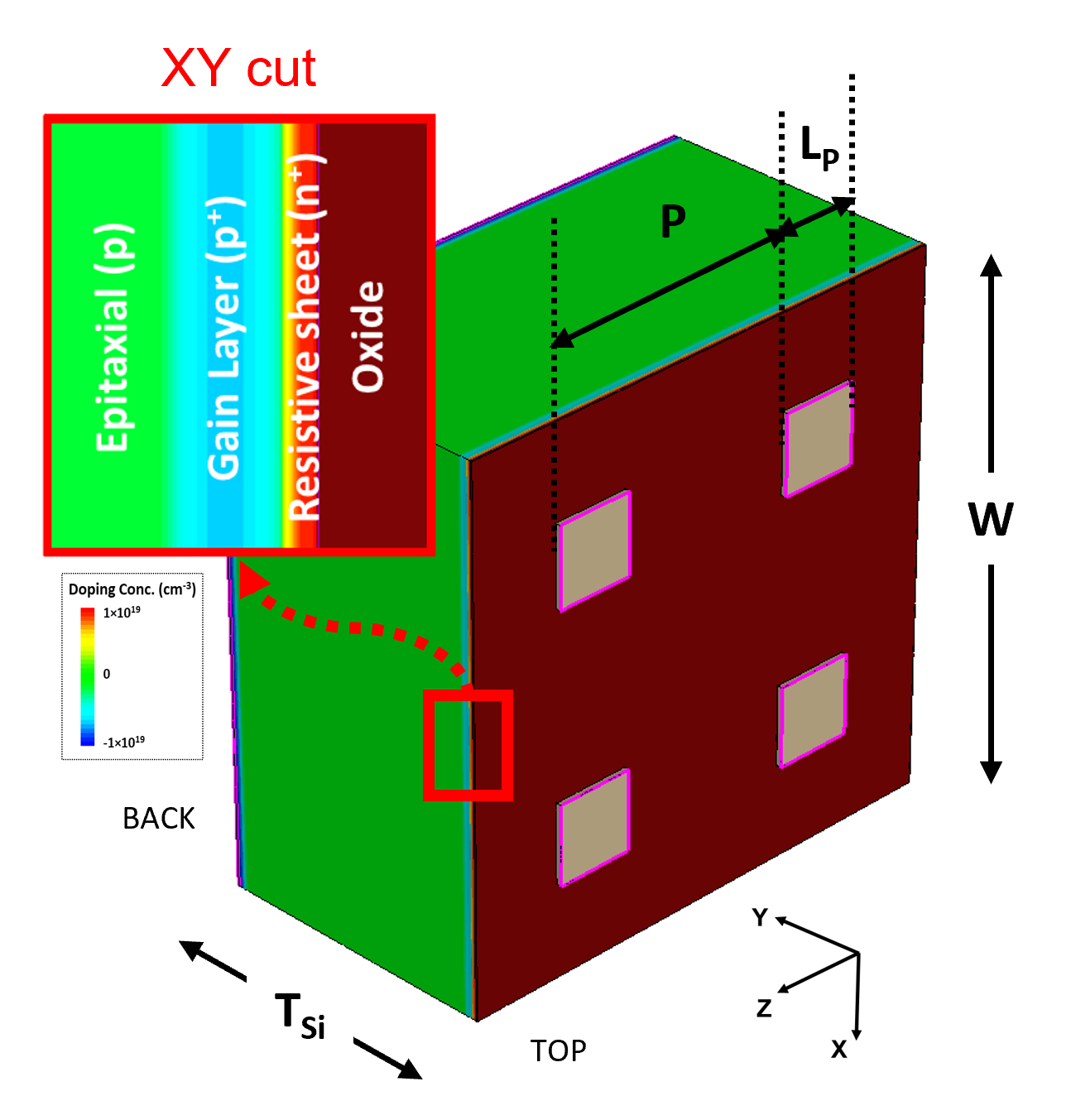}}
\caption{TCAD model of a $50~\text{$\upmu$m}$-pitch four-pad DC-RSD device in the 3D domain. On the top left, an XY cut of the doping profile is reported, where the $n^{+}$-resistive sheet and $p^{+}$-gain layer implants are visible.}
\label{fig31}
\end{figure}

\begin{figure}[t]
\centerline{\includegraphics[width=3.15in]
{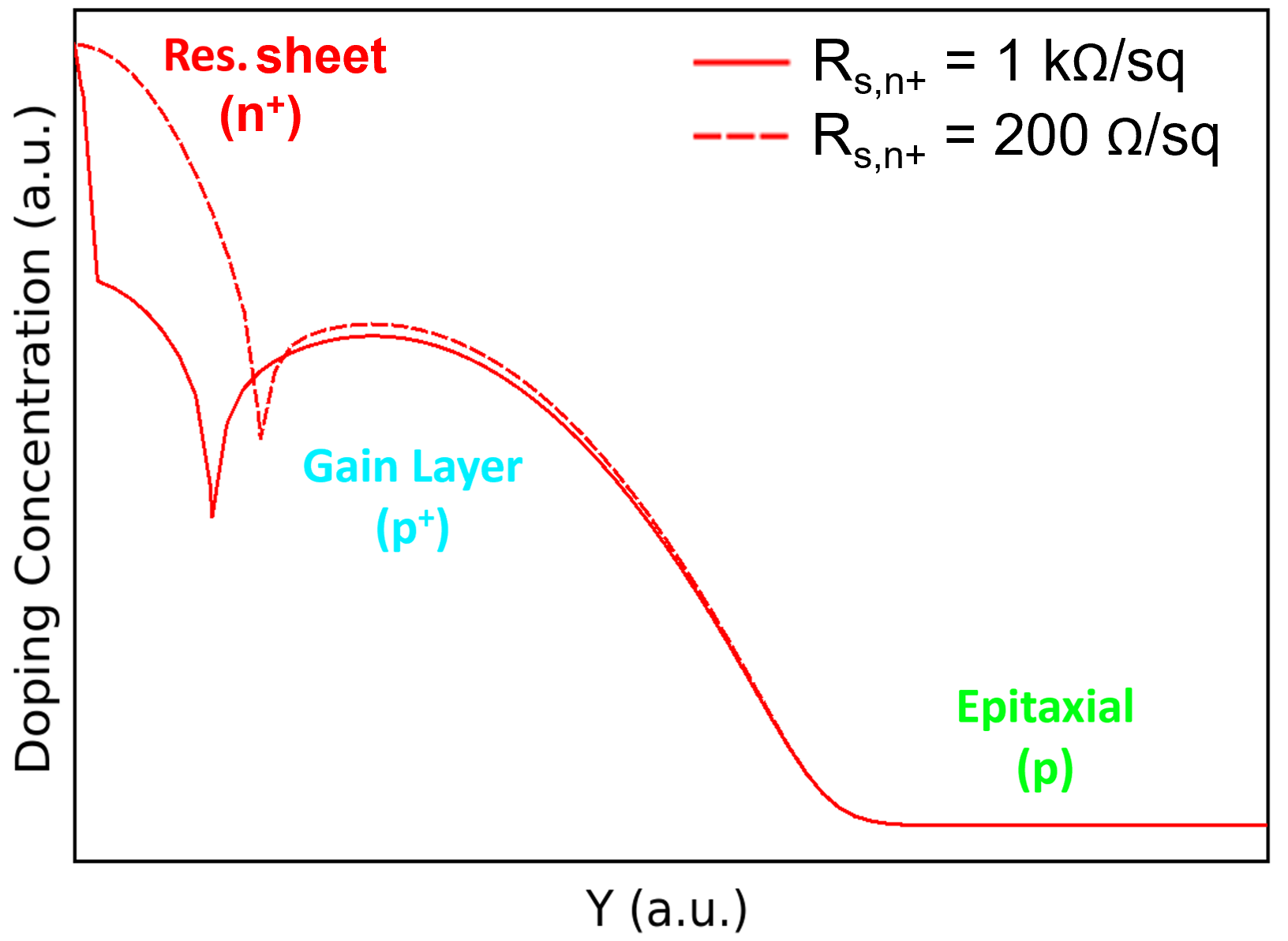}}
\caption{X cut of the doping profile along the first few microns from the surface. Two different $n^{+}$-resistive sheet implants have been considered to study their impact on the sensor electrical behaviour.}
\label{fig32}
\end{figure}

%In TCAD environment \cite{c2}
By using the state-of-the-art Synopsys Sentaurus TCAD suite of tools \cite{d1}, we have modeled a four-pad DC-RSD structure in the 3D domain, considering a proper sensor width ($W$) and thickness ($T_{Si}$), as well as pad length ($L_{P}$) and pitch size (P) - see \figurename~\ref{fig31}. %As represented in the inset of \figurename~\ref{fig31} and in \figurename~\ref{fig32}, the doping profile is the one of an LGAD sensor, for which a $p^{+}$-gain layer has been realized in the very first few microns from the surface. Two different $n^{+}$-resistive sheet implants have been considered to study their impact on the sensor electrical behaviour. In particular, by varying the shape of the $n^{+}$-resistive sheet implant (i.e., the thickness and the doping concentration) as represented in \figurename~\ref{fig32}, we have explored the impact of different sheet resistance values ($R_{s,n^{+}}$) on the steady-state (DC) and transient (TV) behaviour of the sensor.
As represented in the inset of \figurename~\ref{fig31} and in \figurename~\ref{fig32}, a $p^{+}$-gain layer has been realized in the very first few microns from the surface. Furthermore, by varying the shape of the $n^{+}$-resistive sheet implant (i.e., the thickness and the doping concentration), we have explored the impact of different values of the sheet resistance ($R_{s,n^{+}}$) on the steady-state (DC) and transient (TV) behaviour of the sensor.

\begin{figure}[t]
\centerline{\includegraphics[width=3.5in]
% {images/Static_TCADBehavior.png}}
{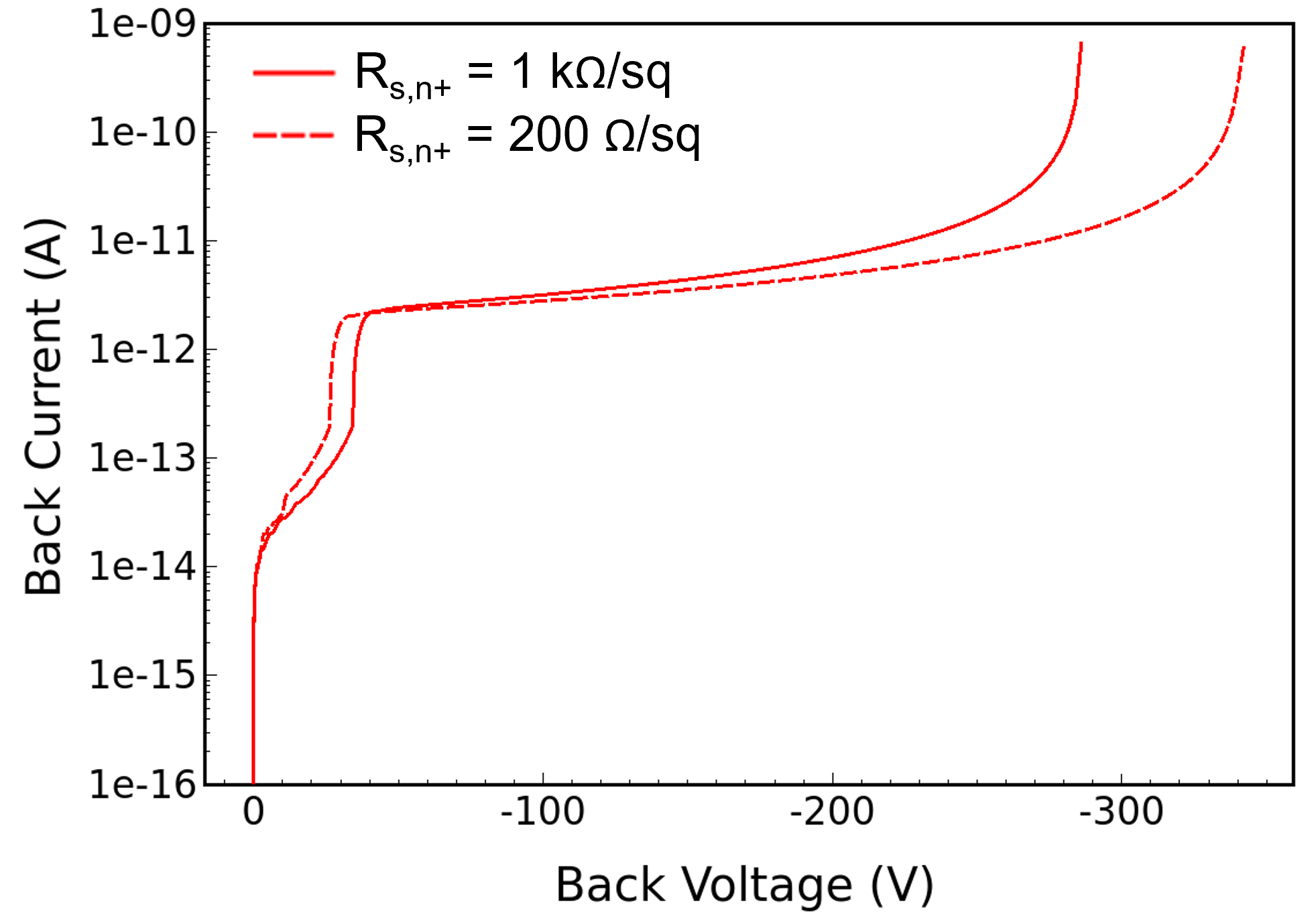}}
\caption{Impact of two different values of the sheet resistance ($200~\text{$\Omega$/sq}$ and $1~\text{k$\Omega$/sq}$) on the DC behavior of a $50~\text{$\upmu$m}$-pitch four-pad DC-RSD detector.}
\label{fig3a}
\end{figure}

In \figurename~\ref{fig3a}, the DC behaviour of the device is described by the current-voltage curves, accounting for two different values of the sheet resistance ($200~\text{$\Omega$/sq}$ and $1~\text{k$\Omega$/sq}$). The reduction of the thickness of the $n^{+}$-resistive sheet leads to an increase in the one of the gain layer implant, and thus a lower breakdown voltage in absolute value. Therefore, when searching for an optimal value of the sheet resistance, it is important to take into account a proper shaping of the gain layer implant to avoid the early breakdown of the device.

\begin{figure}[t]
\centerline{\includegraphics[width=3.5in]
% {images/Transient_TCADBehavior.png}}
{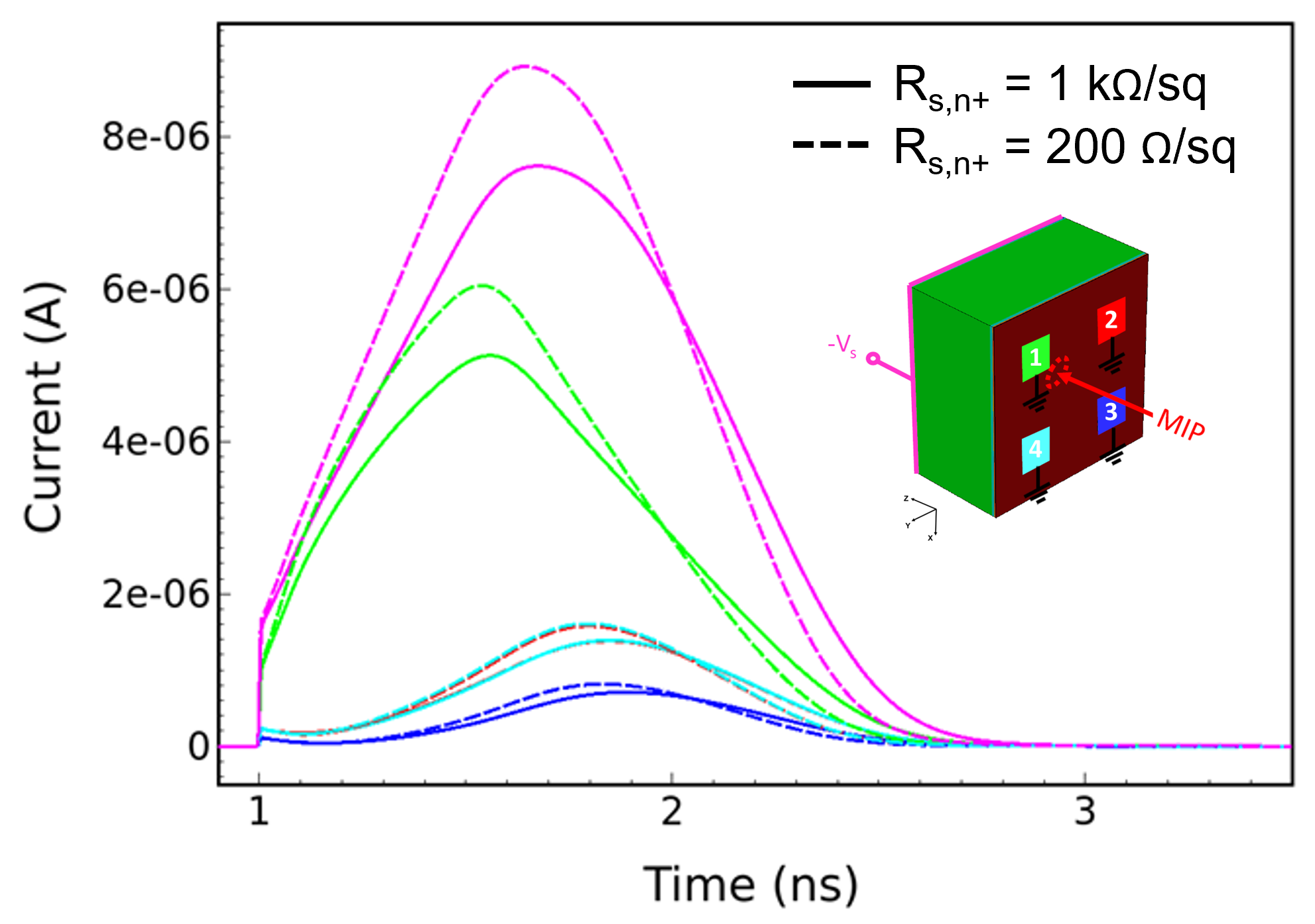}}
\caption{Signals generated by a MIP in a $50~\text{$\upmu$m}$-pitch four-pad DC-RSD detector, biased at 200 V at room temperature. The green, cyan, red and blue curves are the signals read out by the four pads, and the magenta curve is the total signal read out by the contact on the back.}
\label{fig3b}
\end{figure}

\begin{figure}[t]
\centerline{\includegraphics[width=3.3in]
% {images/TCAD_Sim_PosRecon.png}}
{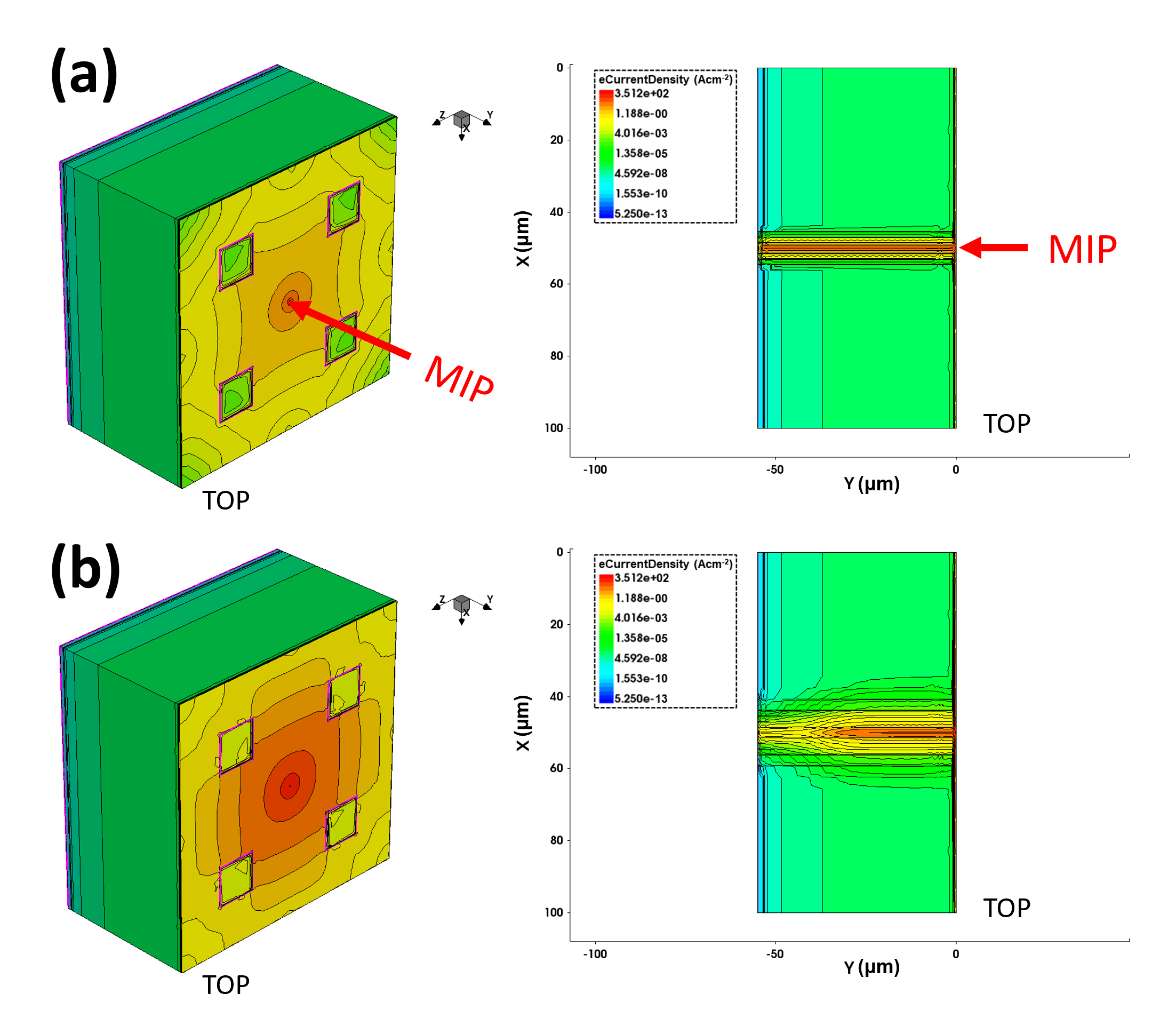}}
\caption{Spatial distribution of electron current over the detector surface (left) and within the bulk (right), immediately after the passage of the MIP (a) and a few hundred picoseconds later (b).}
\label{fig3d}
\end{figure}

In \figurename~\ref{fig3b}, the TV behaviour is described by the current-time curves, accounting for the two values of the sheet resistance mentioned before. The response of the device after the passage of a charged particle (e.g., a minimum ionizing particle - MIP), is represented by the green, red, blue and cyan curves, which are the currents read out by the four pads, and by the magenta curve, which is the current read out on the back contact. %- see \figurename~\ref{fig3b} for the association of the colors between the pads and the read-out signals. 
As already shown by the Spice simulations, the higher the sheet resistance, the lower and wider the read-out signals, worsening the timing capabilities of the detector. Anyway, the sheet resistance value of $1~\text{k$\Omega$/sq}$ ensures better electrical isolation between the pads with respect to $200~\text{$\Omega$/sq}$. By exploiting the 3D modelling feature of the TCAD, it is possible to view the time evolution of the spatial distribution of electron current over the $n^{+}$-resistive sheet and within the bulk, immediately after the passage of the MIP (see \figurename~\ref{fig3d}). 

\begin{figure}[t]
\centerline{\includegraphics[width=3.5in]
{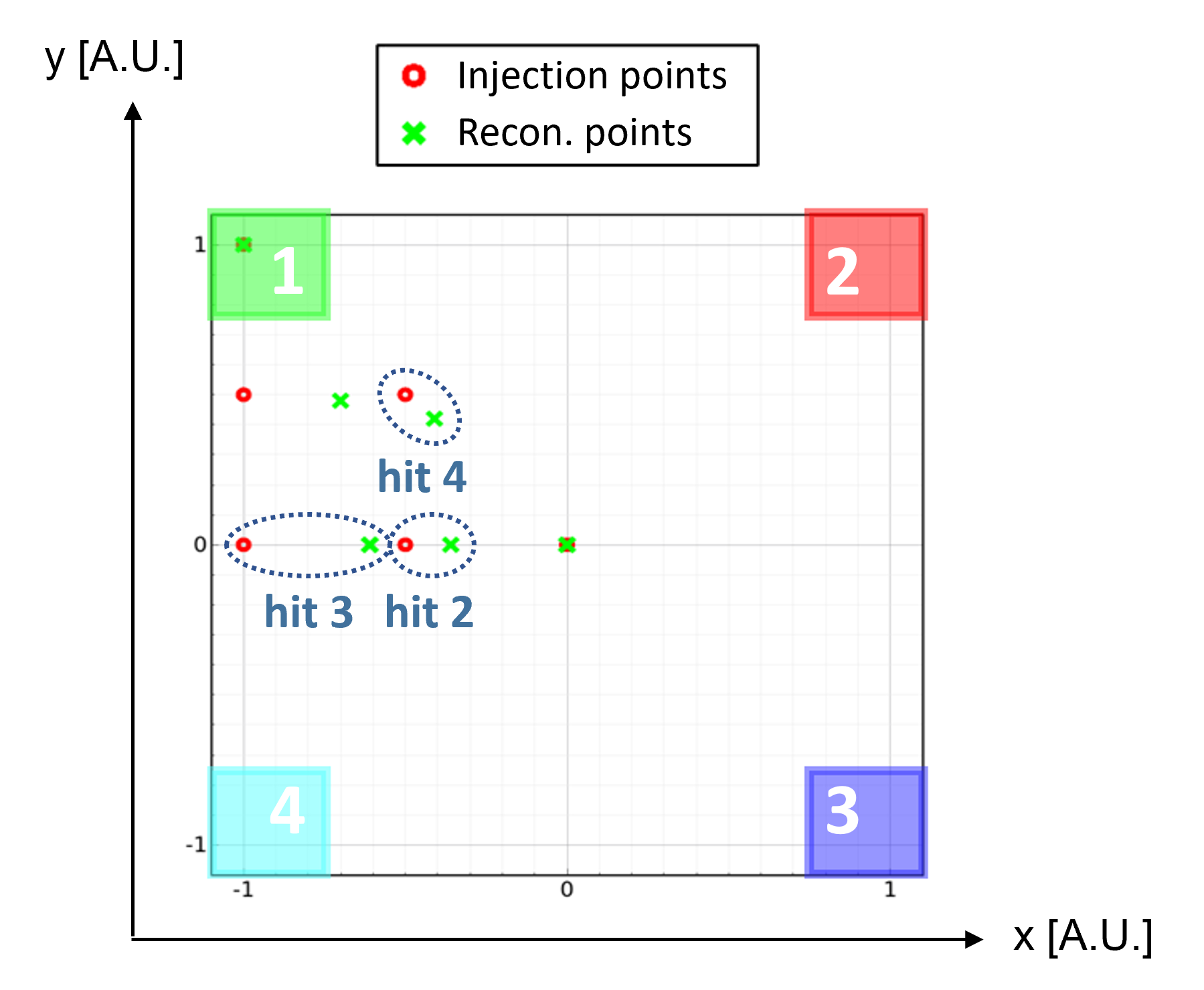}}
\caption{Map of the injected and reconstructed impact positions, which has been obtained by TCAD simulation of a $105~\text{$\upmu$m}$-pitch four-pad DC-RSD detector. The red circles are the injection points, and the green crosses the reconstructed points. Six injection points (hits) have been simulated.} %Note how the latter tend to cluster in the centre, especially at the edges of the sensor.}
\label{fig3c}
\end{figure}

Once the MIP stimulus has been injected into different positions over the detector surface, the generated current spreads over the resistive sheet and it is collected by the four pads. Differently from the Spice simulations, the x-y coordinates of the particle impact positions have been reconstructed by combining the charge extracted from the read-out signals, according to the \textquotedblleft charge imbalance formula\textquotedblright:
\begin{equation}\begin{split}
\label{eq2}
x&=\frac{Q_{2}+Q_{3}-Q_{1}-Q_{4}}{Q_{tot}}\\
y&=\frac{Q_{2}+Q_{1}-Q_{4}-Q_{3}}{Q_{tot}}, 
\end{split}\end{equation}
where $Q_{i}$ is the charge collected by the $i$-th pad, and $Q_{tot}$ is the sum of the charges collected by the four pads. By looking at \figurename~\ref{fig3c}, it is still noticeable how the reconstructed positions do not overlap with the injected ones, thus confirming what has been already observed thanks to the Spice simulations.

%\begin{figure}[t]
%\centerline{\includegraphics[width=3.5in]
%% {images/DCRSD_TCADmodel.png}}
%% {images/DCRSD_TCADmodel_Layout_Doping.png}}
%{images/DCRSD_TCADmodel_Layout_Doping_BIS.png}}
%\caption{(a) TCAD model of a $50~\text{$\upmu$m}$-pitch four-pad DC-RSD device in 3D domain. On the top left, an XY cut of the doping profile is reported, where both the $n^{+}$-resistive sheet and $p^{+}$-gain layer implants are visible. (b) X cut of the doping profile along the first few microns from the surface. Two different $n^{+}$-resistive sheet implants have been considered to study their impact on the sensor electrical behaviour.}
%\label{fig3}
%\end{figure}

%\begin{figure}[t]
%\centerline{\includegraphics[width=3.5in]
%% {images/TCAD_Sim_PosRecon.png}}
%{images/TCAD_Sim_PosRecon_BIS.png}}
%\caption{Map of the injected and reconstructed impact positions that have been obtained by TCAD simulations. The red circles are the injection points, and the green crosses the reconstructed ones.}
%\label{fig3c}
%\end{figure}

\section{Spatial resolution analysis}
\label{sec:resolution}
\begin{figure}[t]
\centerline{\includegraphics[width=3.5in]
{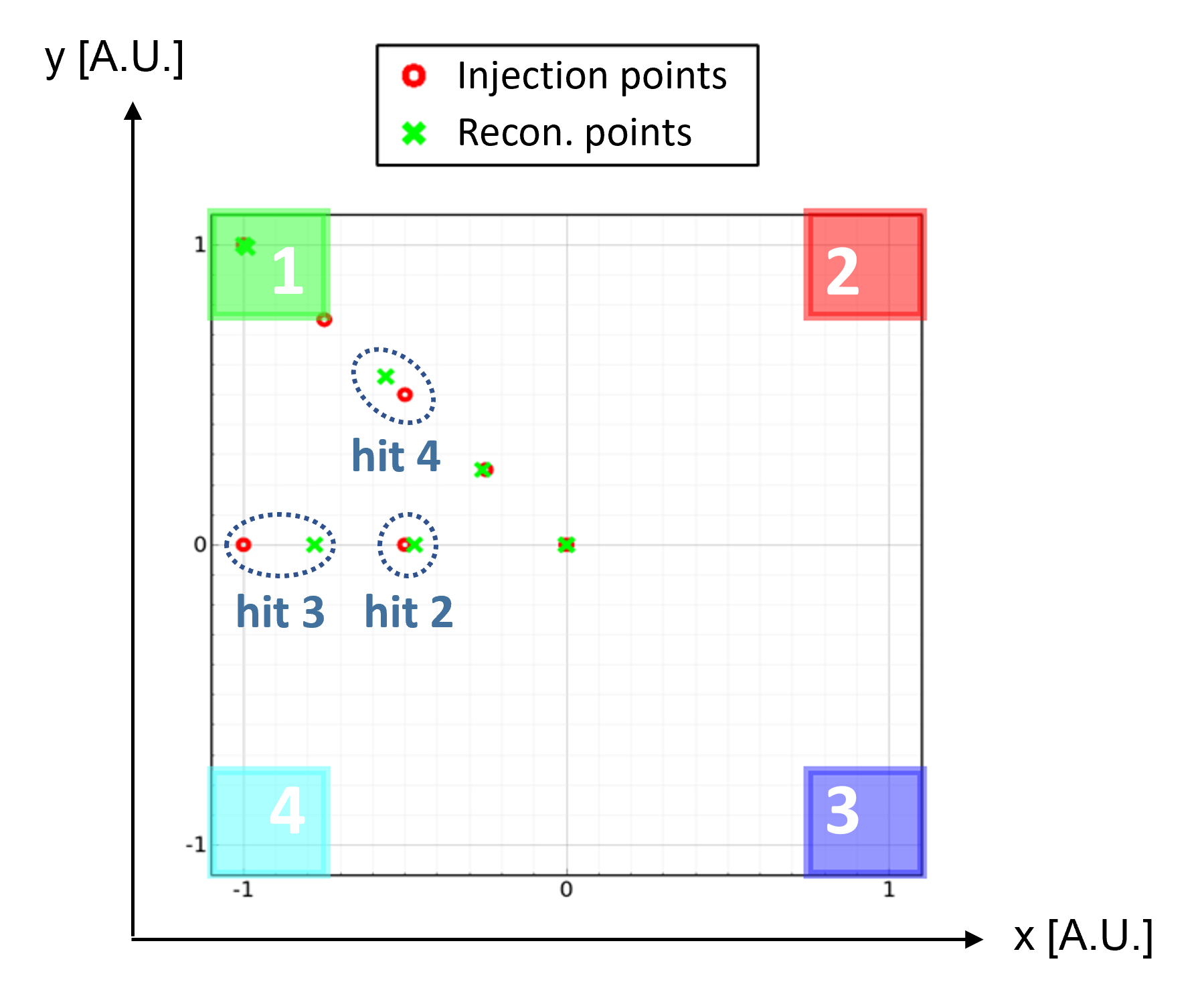}}
\caption{Map of the injected and reconstructed impact positions, which has been obtained by TCAD simulation of a $50~\text{$\upmu$m}$-pitch four-pad DC-RSD detector. The reconstructed points tend to cluster in the centre with a lower degree of distortion with respect to the case in \figurename~\ref{fig3c}.}
\label{fig41}
\end{figure}

By comparing the maps of the reconstructed impact positions obtained with Spice and TCAD simulations (\figurename~\ref{fig2c} and \figurename~\ref{fig3c}, respectively), we have observed that the reconstructed points tend to cluster in the centre, especially those related to the impacts near the edges of the sensor. Thanks to the TCAD simulations, we have also observed that, by fixing the pad length, the smaller the pitch size the lower the degree of the distortion in the reconstruction, as clearly visible by comparing \figurename~\ref{fig3c} and \figurename~\ref{fig41}. 
Such distortion is typical of resistive devices and it is already well documented in literature \cite{d2}. As suggested in \cite{d2}, a proper electrical connection of neighbouring pads provides a low impedance path towards the front-end amplifiers to a higher portion of the charge, allowing for higher collection efficiency, especially for events that took place close to the sensor edges. Moreover, the connection between neighbouring pads allows for insulating one pixel (i.e., a cluster of four pads) from its adjacent, thus improving the signal confinement and the spatial resolution. 

\section{Design and optimization of DC-RSD}
\label{sec:optimization}
\begin{figure}[t]
\centerline{\includegraphics[width=3.45in]
{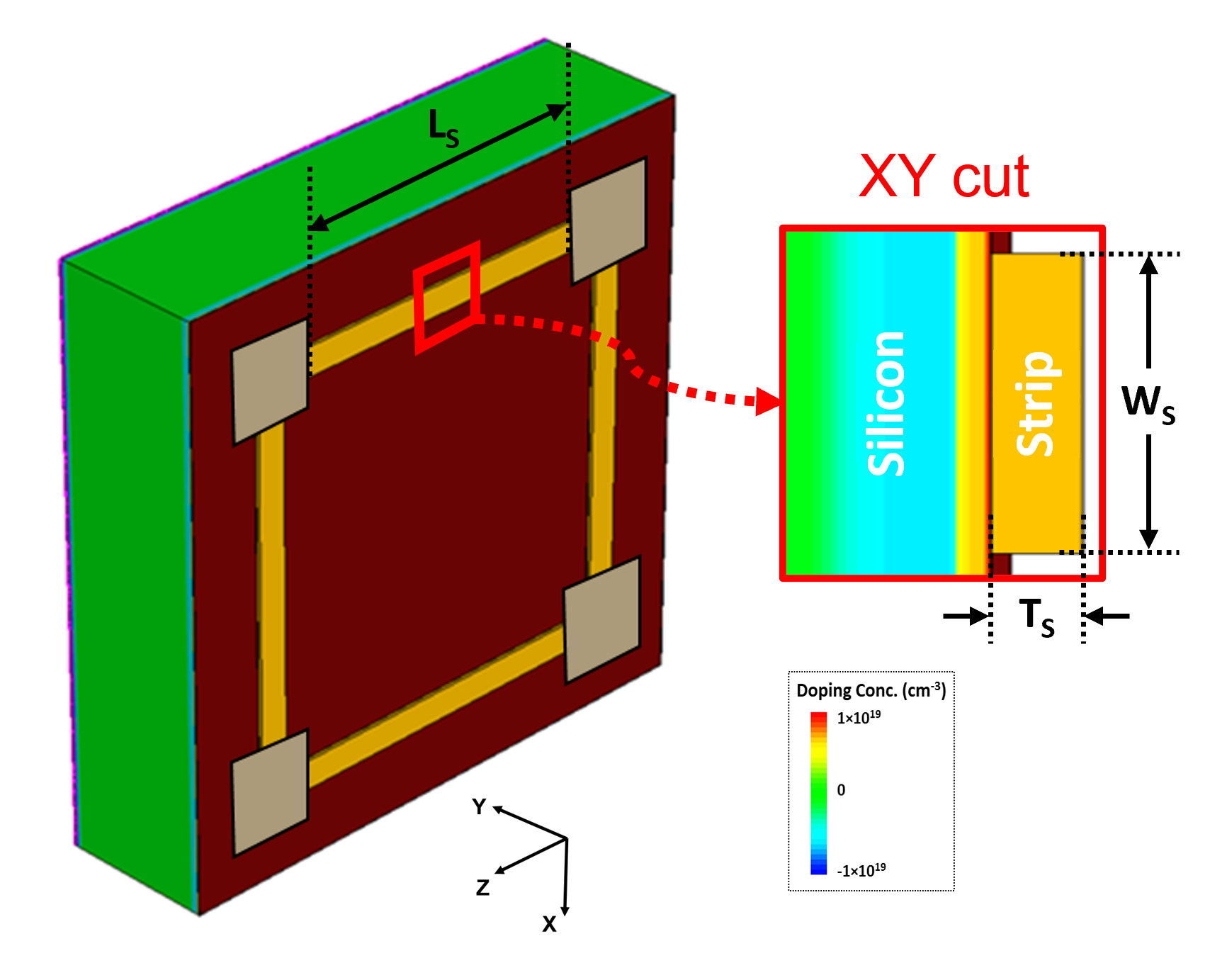}}
\caption{Layout of the strip-connected DC-RSD. The XY cut shows the direct coupling of the strips to the resistive silicon layer. The strip resistance has been tuned by varying the values of the geometrical parameters (length, $L_{S}$, width, $W_{S}$, and thickness, $T_{S}$) and the materials (resistivity) of the strips.}
\label{fig4a}
\end{figure}

According to the suggestions in \cite{d2}, we have modeled within the TCAD environment the strip-connected DC-RSD device represented in \figurename~\ref{fig4a}. As shown in the inset, a direct coupling of the strips to the resistive silicon layer has been realized, as well as a fine-tuning of the strip resistance (i.e., geometry and material) such that it is lower than the resistive sheet, but not enough to short circuit the front-end electronics.

\begin{figure}[t]
\centerline{\includegraphics[width=2.72in]
% {images/TCAD_StripDCRSD_Map.png}}
{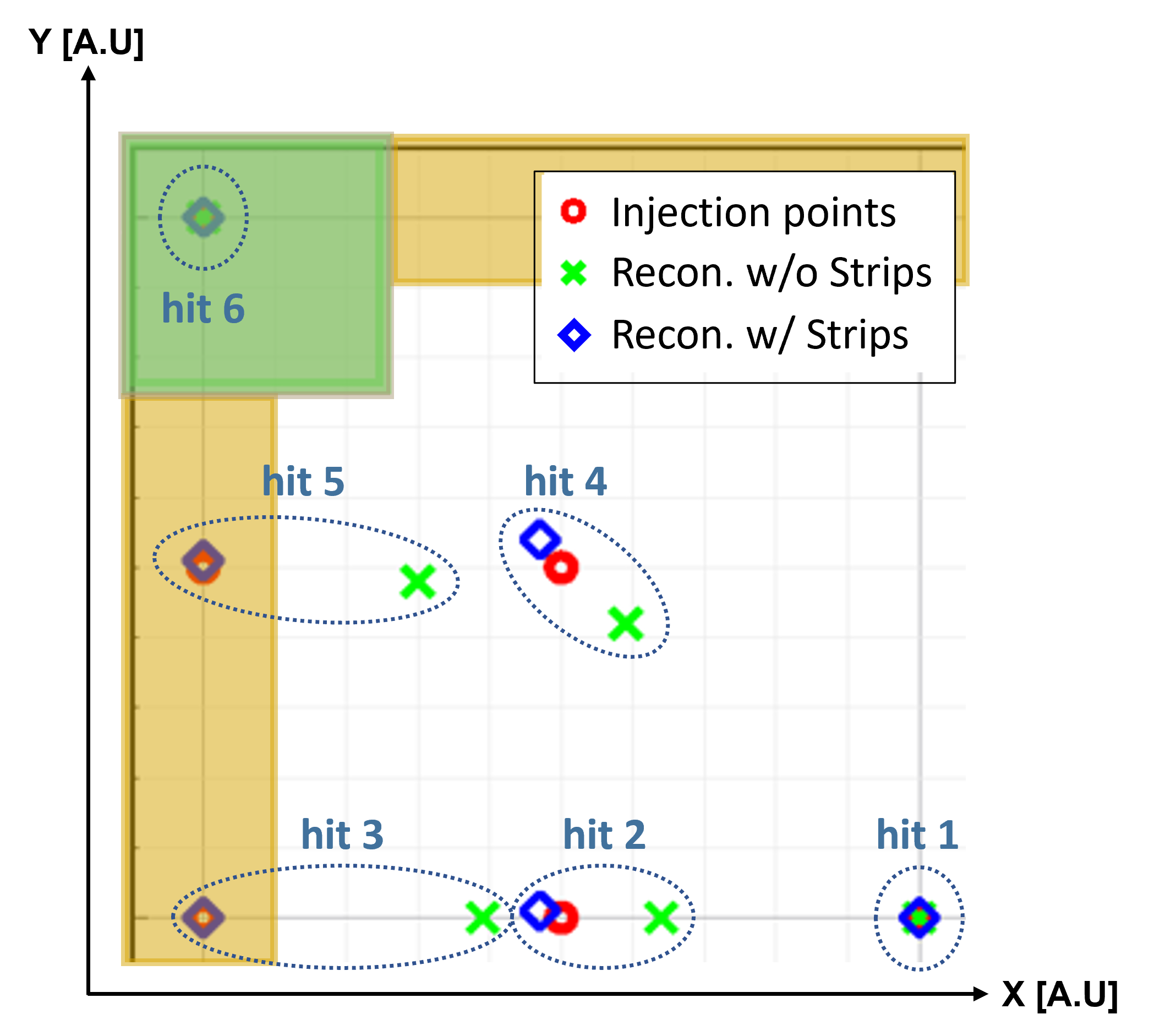}}
\caption{Upper left quadrant of the map of the injected (red circular markers) and reconstructed (blue diamond and green cross markers) particle impact positions. Six injection points (hits) have been simulated in the case of DC-RSD flavor without strips (w/o Strips) and with strips (w/ Strips).}
\label{fig4b}
\end{figure}

\begin{figure}[t]
\centerline{\includegraphics[width=3.6in]
% {images/TCAD_Sim_PosRecon.png}}
{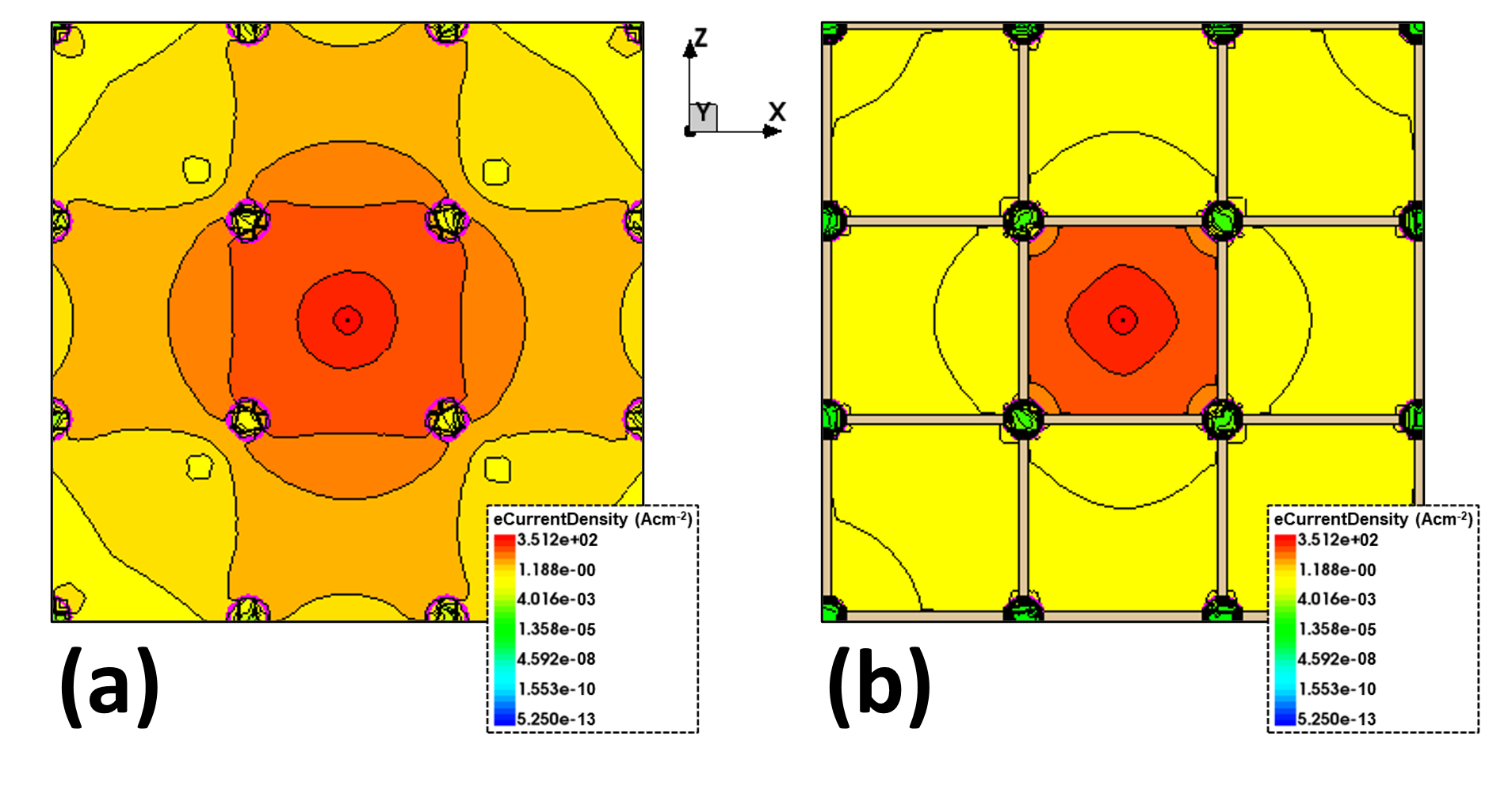}}
\caption{Spatial distribution of electron current over the detector surface of a $20~\text{$\upmu$m}$-pitch sixteen-pad DC-RSD detector without strips (a) and with strips (b), a few hundred picoseconds after the passage of the MIP. The signal confinement within the pixel is better when resistive strips are used to connect neighbouring pads (i.e., the magenta circular dots).}
\label{fig4b_BIS}
\end{figure}

As shown in \figurename~\ref{fig4b}, the distortion is strongly reduced by adding the low-resistive strips between the collecting pads. In the picture, it is represented the upper left quadrant of the map of the injected and reconstructed impact positions. The six red circles are the injection points, and the green crosses and blue diamonds are the reconstructed ones in the case of DC-RSD flavour without strips and with strips, respectively. The accuracy of the position reconstruction improves when inter-pad resistive strips are used, because they help to confine the signal spreading within the pixel (see \figurename~\ref{fig4b_BIS}).

\begin{figure}[t]
\centerline{\includegraphics[width=3.5in]
{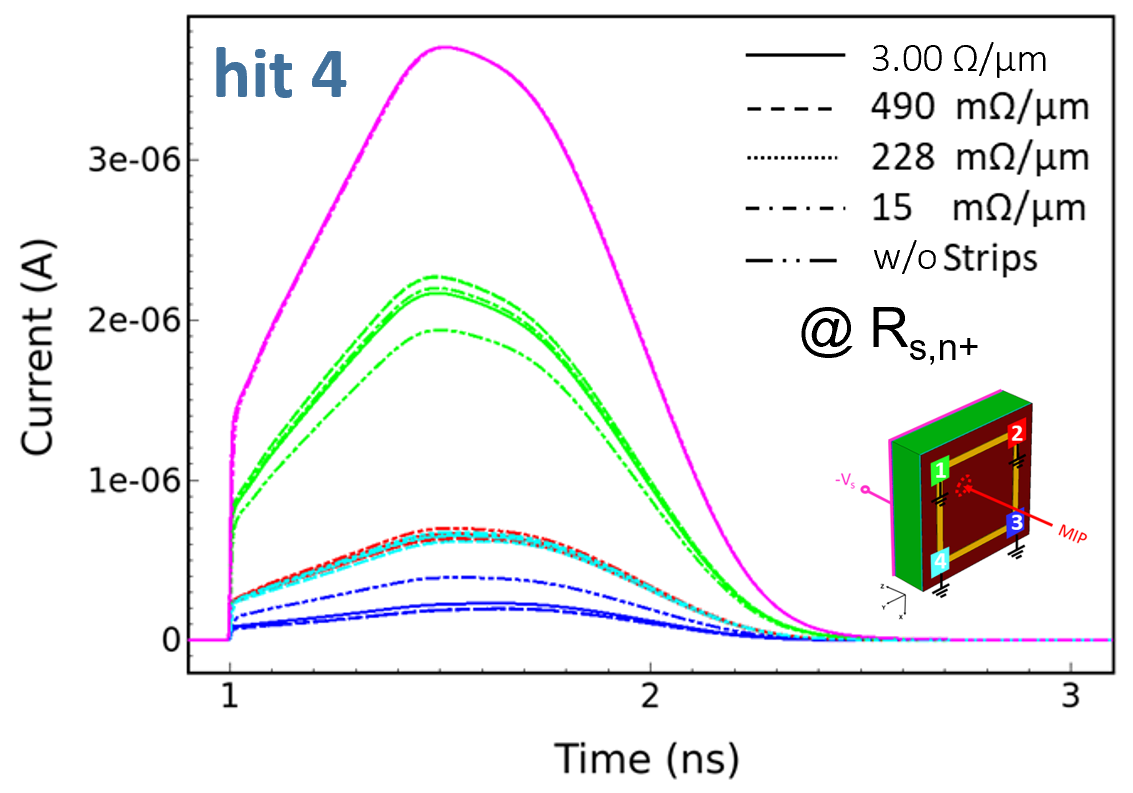}}
\caption{Impact of the strip resistance on the transient behavior of a $105~\text{$\upmu$m}$-pitch four-pad DC-RSD detector. The impact position of the MIP is \textquotedblleft hit 4\textquotedblright.}
\label{fig4c1}
\end{figure}

\begin{figure}[t]
\centerline{\includegraphics[width=3.5in]
{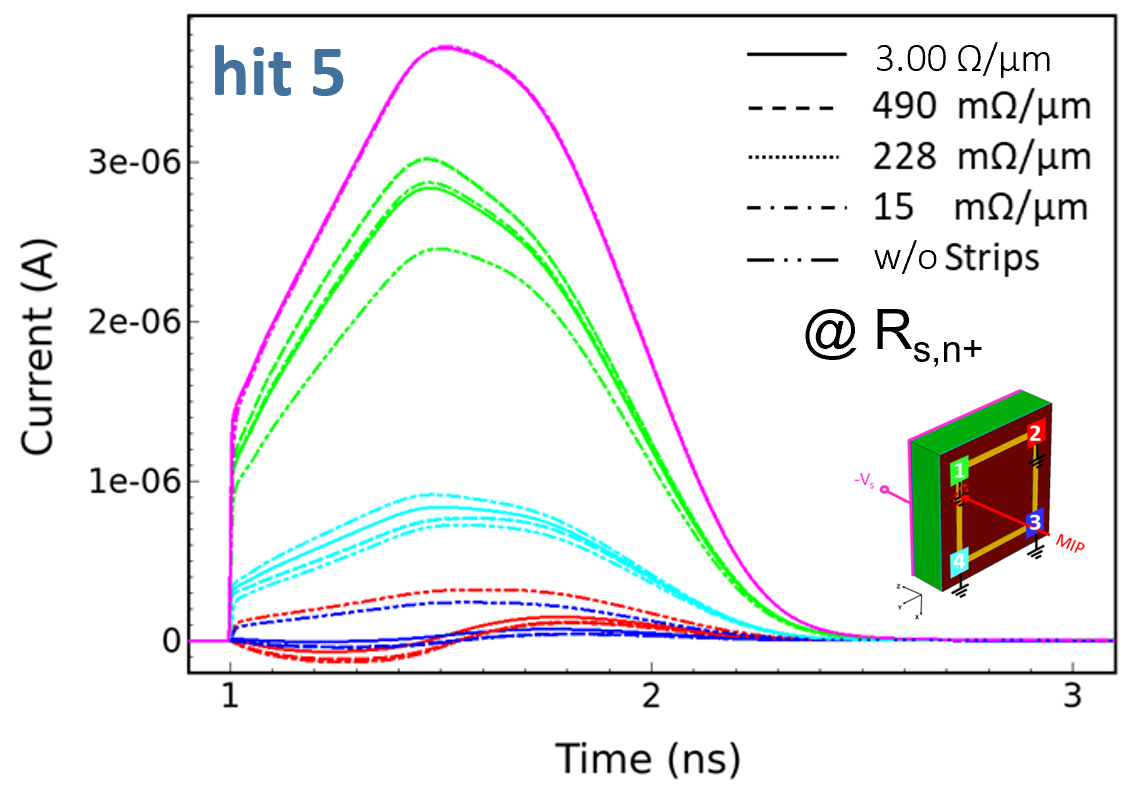}}
\caption{Impact of the strip resistance on the transient behavior of a $105~\text{$\upmu$m}$-pitch four-pad DC-RSD detector. The impact position of the MIP is \textquotedblleft hit 5\textquotedblright.}
\label{fig4c2}
\end{figure}

\begin{figure}[t]
\centerline{\includegraphics[width=2.95in]
% {images/TCADSim_StripRes_Dep.png}}
{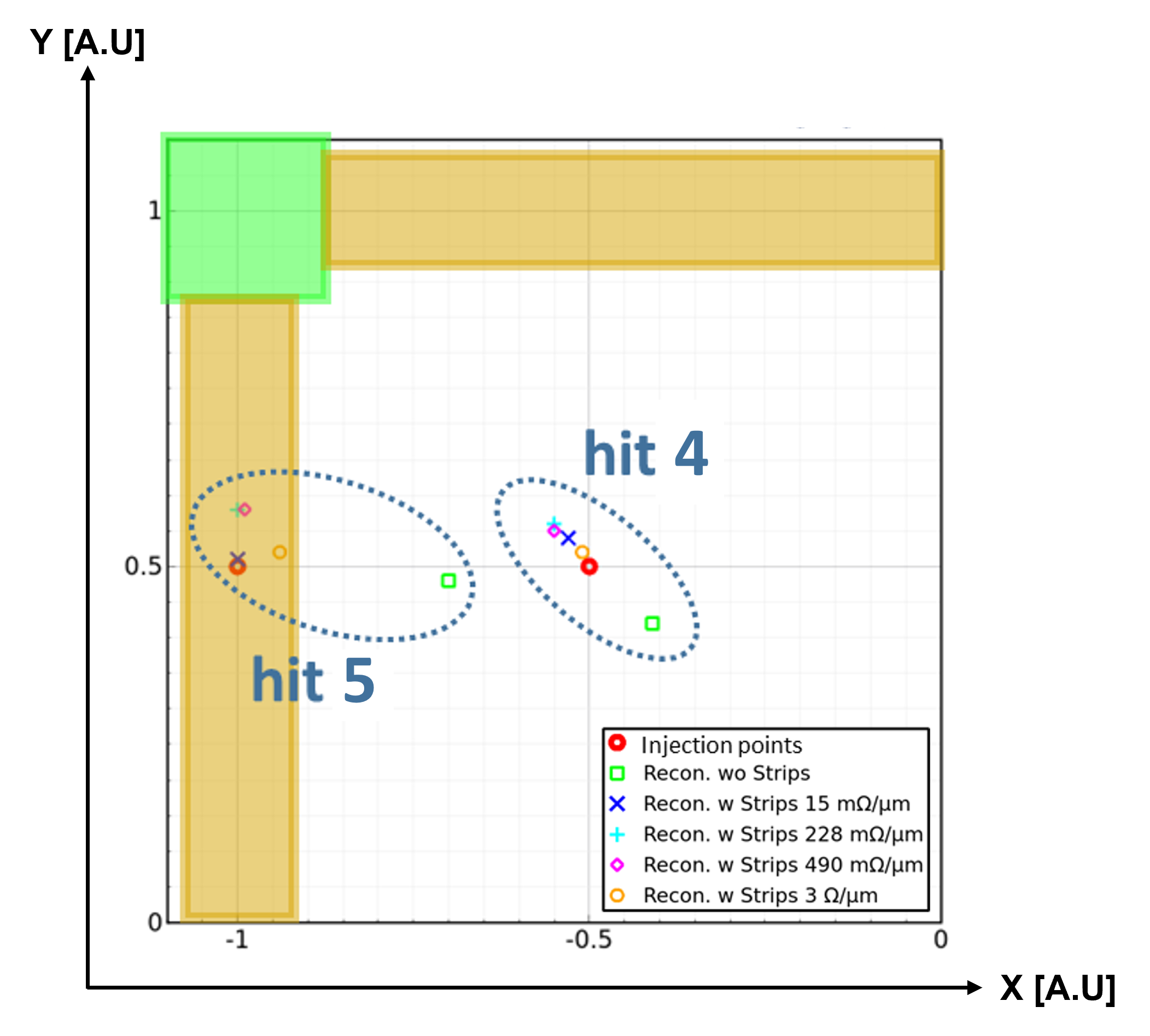}}
\caption{Upper left quadrant of the map of the injected (red circular markers) and reconstructed (all the other markers) particle impact positions for different values of the strip resistance.}
\label{fig4d}
\end{figure}

Thanks to TCAD simulations, we have studied the impact of different values of the strip resistance on the transient behaviour of the device and the reconstruction of the particle impact positions. In particular, when the value of the strip resistance is in the range between $490~\text{m$\Omega$/$\upmu$m}$ and $3~\text{$\Omega$/$\upmu$m}$, the amplitude of the current read out by the pad closest to the injection position tends to be maximized and the one read out by the furthest pads tends to be minimized (see \figurename~\ref{fig4c1} and \figurename~\ref{fig4c2}). Furthermore, when the particle hits a strip, the signal read out by the pads that are on the opposite side becomes bipolar, thus suppressing the contribution of the collected charge from these (see the red and blue lines in \figurename~\ref{fig4c2}). 
All this allows for a better reconstruction of the particle impact positions, as shown in \figurename~\ref{fig4d}, where the red circles are the injection points, the green square markers are the reconstructed points without the use of the low-resistive strips, and all the other markers are the reconstructed points thanks to the presence of the low-resistive strips.

\begin{figure}[t]
\centerline{\includegraphics[width=3.5in]
{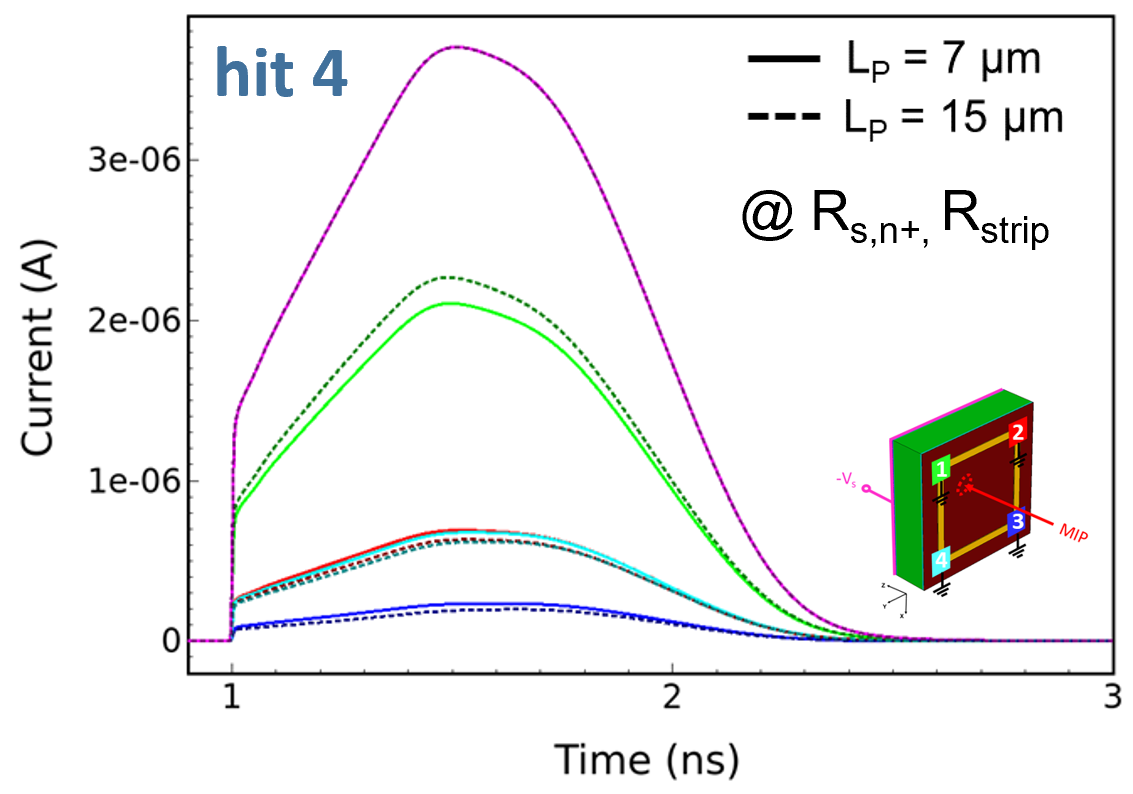}}
\caption{Impact of the pad length on the transient behavior of a $105~\text{$\upmu$m}$-pitch four-pad DC-RSD detector. The impact position of the MIP is \textquotedblleft hit 4\textquotedblright.}
\label{fig4e1}
\end{figure}

\begin{figure}[t]
\centerline{\includegraphics[width=3.5in]
{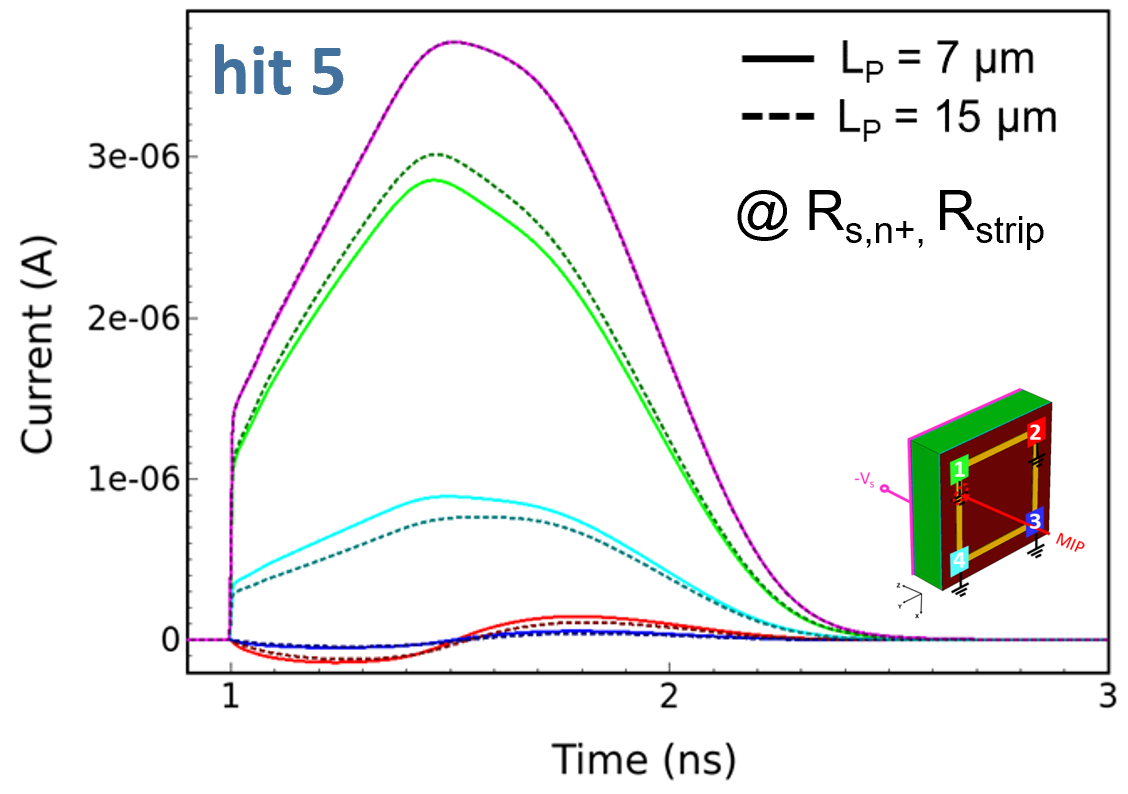}}
\caption{Impact of the pad length on the transient behavior of a $105~\text{$\upmu$m}$-pitch four-pad DC-RSD detector. The impact position of the MIP is \textquotedblleft hit 5\textquotedblright.}
\label{fig4e2}
\end{figure}

\begin{figure}[t]
\centerline{\includegraphics[width=2.95in]
% {images/TCADSim_PadLenght_Dep.png}}
{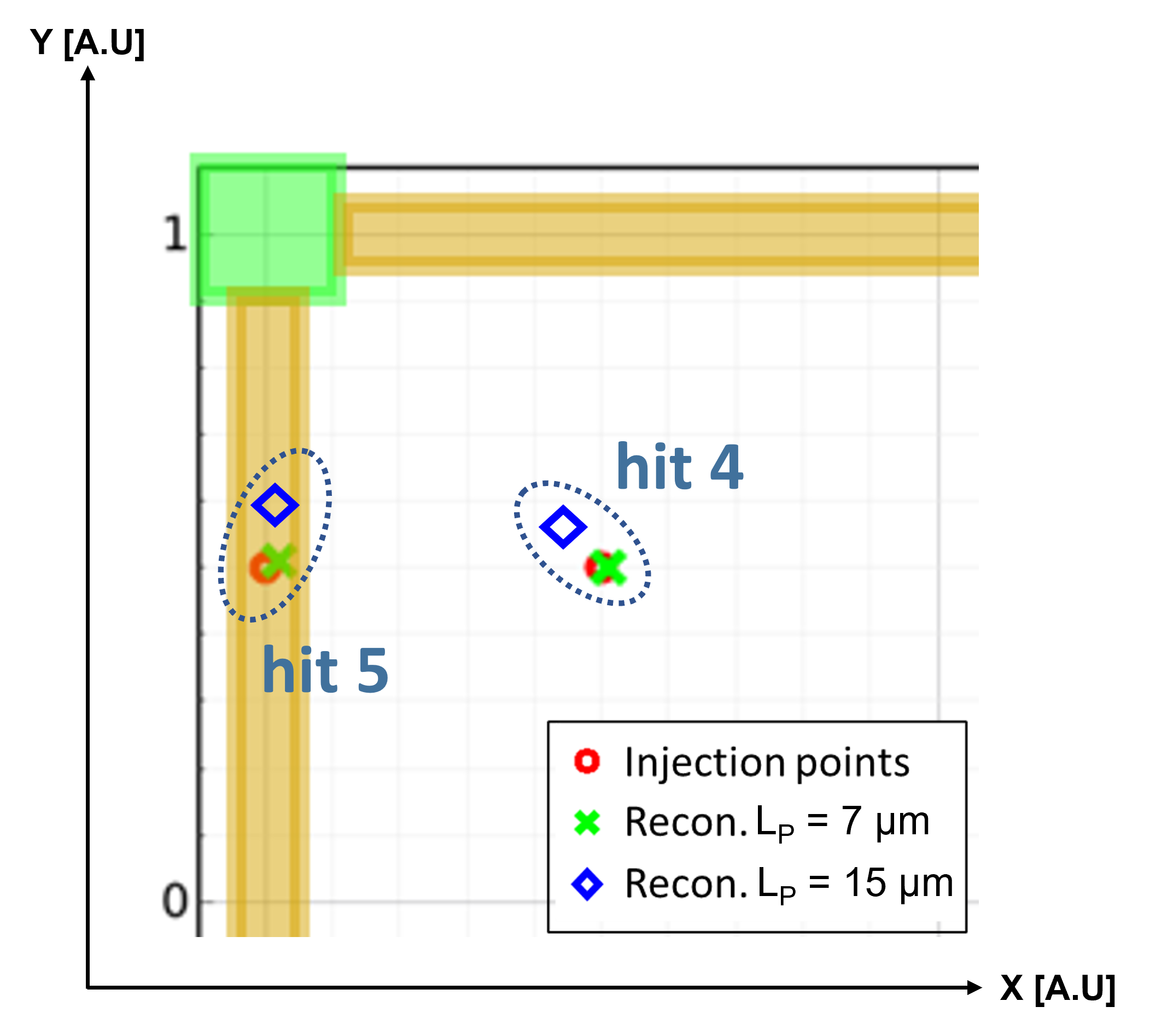}}
\caption{Upper left quadrant of the map of the injected
(red circular markers) and reconstructed particle impact positions for two values of pad length: $7~\text{$\upmu$m}$ (green cross markers), and $15~\text{$\upmu$m}$ (blu diamond markers).}
\label{fig4f}
\end{figure}

Finally, we have also studied the impact of different pad lengths on the transient behaviour and the reconstruction of the particle impact positions. As expected, by reducing the pad length, the amplitude of the read-out currents reduces as well (see the solid lines in \figurename~\ref{fig4e1} and \figurename~\ref{fig4e2}), but this helps to reduce the distortion in the position reconstruction, as shown in \figurename~\ref{fig4f}, where the red circles represent the injection points and the green crosses are the reconstructed points.
%Finally, we have observed that when a particle hits a strip, the signal read out by the pads that are on the opposite side becomes bipolar (see the red and blue lines in \figurename~\ref{fig4c2} and \figurename~\ref{fig4e2}). For this reason, the presence of the strips allows for suppressing the charge collected by these pads, and thus a better reconstruction of the particle impact position (see Fing, hit 5).

%\begin{figure}[t]
%\centerline{\includegraphics[width=3.5in]
%% {images/TCADReconstruction_Distortion.png}}
%{images/TCADReconstruction_Distortion_BIS.png}}
%\caption{Comparison between the maps of the reconstructed impact positions obtained by TCAD simulation of (a) $50~\text{$\upmu$m}$-pitch and (b) $105~\text{$\upmu$m}$-pitch four-pad DC-RSD detectors. The reconstructed points tend to cluster in the centre in both cases, but the larger the pitch size the higher the degree of the distortion, especially at the edges of the sensor.}
%\label{fig4}
%\end{figure}

%\begin{figure}[t]
%\centerline{\includegraphics[width=3.5in]
%{images/TCADReconstruction_Distortion_105umpitch.png}}
%\caption{Map of the reconstructed impact positions obtained by TCAD simulation of a $105~\text{$\upmu$m}$-pitch four-pad DC-RSD detectors. Also in this case the reconstructed points tend to cluster in the centre, but the larger the pitch size the higher the degree of the distortion, especially at the edges of the sensor.}
%\label{fig42}
%\end{figure}

%\begin{figure}[t]
%\centerline{\includegraphics[width=3.5in]
%% {images/TCADSim_StripRes_Dep.png}}
%{images/TCADSim_StripRes_Dep_It.png}}
%\caption{Impact of the strip resistance on the transient behavior of a $105~\text{$\upmu$m}$-pitch four-pad DC-RSD detectors for two different particle impact positions: (a) hit 3, and (b) hit 8.}
%\label{fig4c}
%\end{figure}

%\begin{figure}[t]
%\centerline{\includegraphics[width=3.5in]
%% {images/TCADSim_PadLenght_Dep.png}}
%{images/TCADSim_PadLenght_Dep_It.png}}
%\caption{Impact of the pad length on the transient behavior of a $105~\text{$\upmu$m}$-pitch four-pad DC-RSD detectors for two different particle impact positions: (a) hit 3, and (b) hit 8.}
%\label{fig4e}
%\end{figure}

\section{Conclusion}
In this work, we propose a novel evolution of the LGAD-based resistive silicon detector design: the DC-coupled RSD. By using a DC-coupled read-out, it is possible to overcome the well-known issues that have been observed for the AC-coupled RSDs (e.g. baseline fluctuation, long-tail bipolar signals). 

A strategy for the simulation of DC-RSD devices has been presented as well, which is based on a two-step procedure: 

\emph{(i)} modeling of the full sensor module by means of an equivalent lumped-element electrical circuit in Spice environment, and simulation of the output waveforms by injecting into the circuital nodes a test input signal. This allows to extract the most important design parameters (e.g., sheet resistance, pad size and pitch) with very short simulation times;

\emph{(ii)} full 3D TCAD simulation to optimize the design parameters by evaluating with high accuracy their impact on the electrical behaviour and the response of the detector to a stimulus (e.g., a MIP) directly from its contacts, overcoming the bandwidth limitation imposed by the read-out circuitry.

%Thanks to the hybrid approach that combines the use of TCAD and Spice simulation tools, it has been demonstrated that the key features of the RSDs, i.e. signal spreading and $100\%$ fill factor, are well preserved in the DC-RSD design. Moreover, according to the simulations, the accuracy of the reconstruction of the particle impact positions improves by using resistive strips between the read-out electrodes. 

Thanks to the hybrid approach that combines the use of TCAD and Spice simulation tools, it has been demonstrated that the accuracy of the reconstruction of the particle impact positions improves by using low-resistive strips between the read-out electrodes. A quantitative evaluation of the effects of the technology realization of the resistive strips (i.e., doping, geometry and material) has been pursued, thus giving in advance information and guidelines for their fabrication.

A first batch of DC-RSD devices is planned in the next few months and an extensive campaign of measurements has been foreseen on both not irradiated and irradiated structures. By comparing the experimental data with a new batch of TCAD simulations that properly take into account the radiation damage effects, we will have new guidelines for designing future production of radiation-resistant DC-RSD sensors.

\section*{Acknowledgment}

This project has received funding from the European Union's Horizon 2020 research and innovation programme under GA No 101004761 and from the Italian MIUR PRIN under GA No 2017L2XKTJ.

\end{document}